\documentclass[english,prl , twocolumn]{revtex4}
\usepackage{ae,aecompl}
\usepackage{helvet}
\usepackage[T1]{fontenc}
\usepackage[latin9]{inputenc}
\usepackage{color}
\usepackage{float}
\usepackage{textcomp}
\usepackage{amstext}
\usepackage{amssymb}
\usepackage{graphicx}

\makeatletter

\DeclareFontEncoding{LGR}{}{}
\DeclareTextSymbol{\~}{LGR}{126}

\providecommand{\tabularnewline}{\\}

\@ifundefined{textcolor}{}
{%
 \definecolor{BLACK}{gray}{0}
 \definecolor{WHITE}{gray}{1}
 \definecolor{RED}{rgb}{1,0,0}
 \definecolor{GREEN}{rgb}{0,1,0}
 \definecolor{BLUE}{rgb}{0,0,1}
 \definecolor{CYAN}{cmyk}{1,0,0,0}
 \definecolor{MAGENTA}{cmyk}{0,1,0,0}
 \definecolor{YELLOW}{cmyk}{0,0,1,0}
 }

\makeatother

\usepackage{babel}
\begin{document}

\title{Protected Josephson Rhombi Chains}

\author{Matthew T. Bell$^{1}$, Joshua Paramanandam$^{1}$, Lev B. Ioffe$^{1,2}$,
and Michael E. Gershenson$^{1}$ }

\address{$^{1}$Department of Physics and Astronomy, Rutgers University, Piscataway,
New Jersey 08854}

\address{$^{2}$LPTHE, CNRS UMR 7589, 4 place Jussieu, 75252 Paris, France }
\begin{abstract}
We have studied the low-energy excitations in a minimalistic protected
Josephson circuit which contains two basic elements (rhombi) characterized
by the $\pi$ periodicity of the Josephson energy. Novel design of
these elements, which reduces their sensitivity to the offset charge
fluctuations, has been employed. We have observed that the life time
{\normalsize $T_{1}$} of the first excited state of this quantum
circuit in the protected regime is increased up to 70{\normalsize $\mu s$},
a factor of $\sim100$ longer than that in the unprotected state.
The quality factor {\normalsize $\omega_{01}T_{1}$} of this qubit
exceeds {\normalsize $10^{6}$}. Our results are in agreement with
theoretical expectations; they demonstrate the feasibility of symmetry
protection in the rhombi-based qubits fabricated with existing technology. 
\end{abstract}
\maketitle
\textcolor{black}{Quantum computing requires the development of quantum
bits (qubits) with a long coherence time and the ability to manipulate
them in a fault tolerant manner (see, e.g. \cite{Knill05} and references
therein). Both goals can be achieved by the realization of a protected
logical qubit formed by a collective state of an array of faulty qubits
\cite{Kitaev20032,ioffe2002a,Ioffe2002b,doucot2002a,doucot2003,doucot2005}.
The building block (i.e. the faulty qubit) of the array is the Josephson
element with an effective Josephson energy $E(\phi)=-E_{2}\cos(2\phi)$,
which is $\pi$ - periodic in the phase difference $\phi$ across
the element. In contrast to the conventional Josephson junctions with
$E(\phi)=-E_{1}\cos(\phi)$, this element supports the coherent transport
of pairs of Copper pairs (the \textquotedblleft{}4e\textquotedblright{}
transport), whereas single Cooper pairs are localized and the {}``2e''
transport is blocked \cite{doucot2002a,protopopov2004,protopopov2006}.
Though this proposal has attracted considerable theoretical attention
\cite{brooks2013}, the experimental realization of a protected qubit
was lacking. }

In this Letter we make an essential step towards building a protected
Josephson qubit by fabricating the simplest protected circuit and
demonstrating that the first excited state of the circuit is protected
from energy relaxation. 

The idea of protection is illustrated in Fig.1. Let us consider the
simplest chain of two $\cos(2\phi)$ elements. They share the central
superconducting island whose charge is controlled by the gate. The
Hamiltonian of this quantum circuit can be written as 
\begin{equation}
H=-2E_{2}\cos(2\phi)+E_{C}(n-n_{g})^{2}\label{eq:H}
\end{equation}
 where the energy $E_{2}$ describes the Josephson coupling of the
central superconducting island to the current leads, $E_{C}$ is the
effective charging energy of the island, $n$ is the number of Cooper
pairs on the island, $n_{g}$ is the charge induced on the island
by the gate. The parity of $n$ is preserved if the transfer of single
Cooper pairs is blocked ($E_{1}=0$). In this case the states of the
system can be characterized by the quantum number $\aleph=n\, mod(2)$.
The low energy states corresponding to number $\aleph=0,1$ are shown
in Fig.1. The energy $E_{2}$ plays the role of the kinetic term that
controls the \textquotedblleft{}spread\textquotedblright{} of the
wave functions along the $n$ axis. Provided $E_{2}\gg E_{C}$, the
number of components with different $n$ in these discrete Gaussian
wavefunctions is large: $\left\langle n^{2}\right\rangle =\sqrt{E_{2}/E_{C}}\gg1$,
and the energy difference between the two states, $E_{01}=E_{\left|1\right\rangle }-E_{\left|0\right\rangle }$,
is exponentially small (see \cite{doucot2012} and Supplementary Material
1):
\begin{equation}
E_{01}=4A\left(g\right)g^{1/2}\exp\left(-g\right)\cos(\pi n_{g})\:\omega_{P}.\label{eq:E01}
\end{equation}
Here $g=4\sqrt{E_{2}/E_{C}}$, $\omega_{P}=4\sqrt{E_{2}E_{C}}/\hbar$
is the plasma frequency, $A\left(g\right)\sim1$ (Supplementary Material
1). Furthermore, these states cannot be distinguished by the noise
operators, and, thus, the decay and dephasing rates are both reduced
by the same large factor $\exp(g)$. 

In real circuits, the 2e processes are not completely suppressed,
and a non-zero amplitude $E_{1}$ mixes the odd and even components
(Fig.1) and increases $E_{01}$. For a small amplitude $E_{1}$ the
decay of the first excited state is due to the mixture of $\aleph=0,1$
states and is suppressed by the factor $(E_{01}/E_{1})^{2}>1$ in
addition to the suppression by the factor $\exp(g)$ that is common
to decay and dephasing. Thus, for the coherence protection two conditions
are required: $E_{2}\gg E_{1}$ (i.e. slow energy relaxation) and
$E_{2}\gg E_{C}$ (i.e. small dephasing rate). 

\begin{figure}
\includegraphics[scale=0.35]{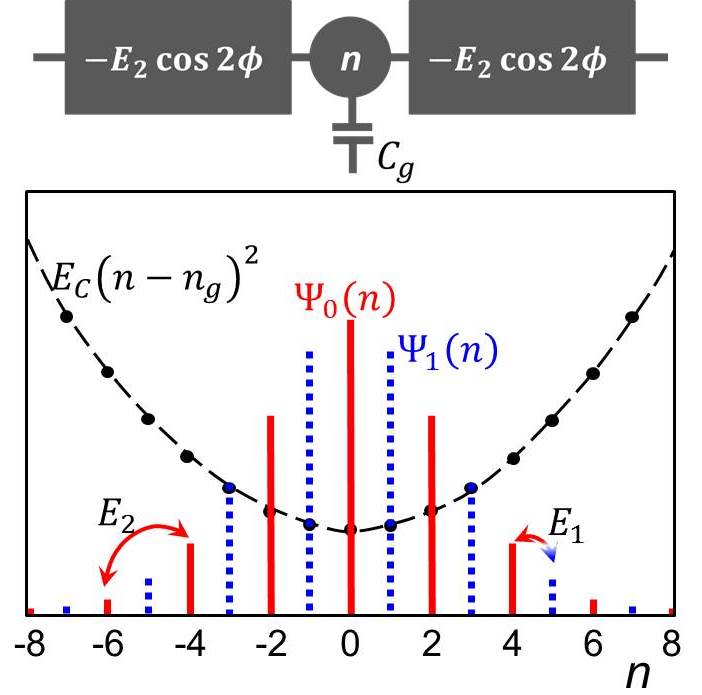}\caption{Top panel: The chain of two $cos(2\varphi)$ element;
the charge of the central (common to both rhombi) island is controlled
by the gate voltage. Bottom panel: Two lowest-energy wavefunctions
in the discrete harmonic potential shown for $n_{g}=0$.}

\end{figure}

The simplest $\cos(2\phi)$ Josephson elements is represented by the
Josephson rhombus: a superconducting loop interrupted by four identical
Josephson junctions \cite{doucot2002a,protopopov2004,protopopov2006,gladchenko,pop2008}
(Fig.2a). When the rhombus is threaded by the magnetic flux $\Phi_{R}=\Phi_{0}/2$
($\Phi_{0}$ is the flux quantum), its effective Josephson energy
$E_{R}=-E_{2R}\cos(2\phi)$ becomes $\pi$-periodic in the phase difference
across the rhombus, $\phi$. In line with theoretical predictions,
recent experiments \cite{gladchenko} have demonstrated that the properly
designed small rhombi arrays can support a non-zero 4e supercurrent
in the regime when the 2e supercurrent vanishes. 

\begin{figure}
\includegraphics[scale=1.1]{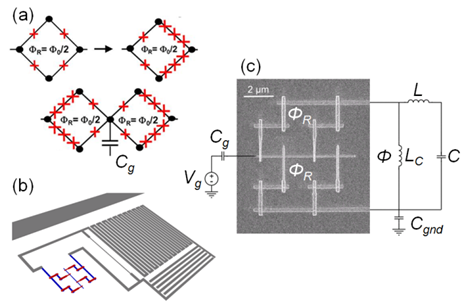}\caption{Panel (a): The $\cos(2\varphi)$ Josephson element
(the Josephson rhombus) and its improved version which is less sensitive
to the random offset charges on the upper and lower islands \cite{doucot2012}.
The chain contains two rhombi; the charge of the central (common to
both rhombi) island is controlled by the gate voltage applied between
the central conductor of the microstrip line and the ground. Panel
(b): The on-chip circuit layout of the device inductively coupled
to the microstrip transmission line. Panel (c): The micrograph of
the two-rhombi chain coupled to the read-out $LC$ resonator via the
kinetic inductance $L_{C}$ of a narrow superconducting wire. The
magnetic flux trough each rhombus is $\Phi_{R}$; the flux in the
phase loop is $\Phi$. }
\end{figure}

In the current work, we have implemented the two-rhombi chain with
an improved design of individual rhombi proposed in Ref. \cite{doucot2012}.
The key requirement for the protection is the cancellation of $E_{1}$
due to the destructive interference between the Cooper pair transfer
amplitudes along the upper and lower branches of a rhombus. This cancellation
is difficult to achieve in the quantum regime where the amplitudes
depend on the uncontrolled offset charges on the side islands. This
dependence is due to the Aharonov-Casher effect (see e.g. \cite{pop2012})
and the high probability of phase slips across the two small junctions
in each arm of a rhombus. In the improved rhombus design \cite{doucot2012}
one junction in each branch is replaced by a short chain of larger
junctions (Fig.2a). The phase slips across the larger junctions are
suppressed due to a large $E_{JL}/E_{CL}$ ratio ($50$ for the studied
devices). As a result, an improved rhombus becomes insensitive to
the offset charges on all islands except for the central island shared
by both rhombi. Below we refer to the characteristic energies of the
smaller and larger junctions as $E_{JS}$,$E_{CS}$ and $E_{JL}$,$E_{CL}$
respectively. An optimal operation of the rhombus is realized for
$E_{JL}=mE_{JS}$ and $E_{CL}=E_{CS}/m$, where $m$ is the number
of larger junctions in the chain ($m=3$ for the studied devices). 

The chains, the readout circuits, and the microwave (MW) transmission
line were fabricated using multi-angle electron-beam deposition of
Al films through a lift-off mask (for fabrication details see Refs.
\cite{bell2012,bell2012a}). The in-plane dimensions of the small
and large junctions were $0.14\times0.13\,\mu m^{2}$ and $0.25\times0.25\,\mu m^{2}$,
respectively. The ratio $E_{JS}/E_{CS}\approx3-5$ was chosen to realize
the resonance frequency of the $\left|0\right\rangle \rightarrow\left|1\right\rangle $
transition, $f_{01}$, within the ($1-10$) GHz range. Below we show
the data for one representative device with $E_{JL}=190GHz$, $E_{CL}=4GHz$,
$E_{JS}=60GHz$, and $E_{CL}=18.6GHz$ (throughout the paper all energies
are given in the frequency units, $20GHz\approx1K$). 

The readout lumped-element $LC$ resonator was formed by the meandered
2-${\mu}$m-wide Al wire with the kinetic inductance $L=3$ nH
and an interdigitated capacitor $C=100$ fF. The resonator was coupled
to the chain via a narrow superconducting wire with a kinetic inductance
of $L_{C}=0.4nH$. The chain and the wire formed a superconducting
loop; the flux $\Phi$ of the external magnetic
field in this phase loop controlled the phase difference across the
chain $\varphi\equiv2\pi\Phi/\Phi_{0}$
(Fig.2c). Because the phase loop area ($1,140\mu m^{2}$) was much
greater than the rhombus area ($13.5\mu m^{2}$), the phase across
the chain could be varied at an approximately constant value of $\Phi_{R}$.
Several devices with systematically varied values of $E_{JS}$ and
$E_{CS}$ were fabricated on the same chip and inductively coupled
to the same microstrip line (Fig.2b). The devices could be individually
addressed due to different resonance frequencies of the $LC$ resonators.
All measurements have been performed at $T$=20 mK.

In the experiment, the microwaves traveled along the microstrip line
inductively coupled to the device. The microwaves at the frequency
$\omega_{1}$ probed the $LC$ resonance. The
microwaves at the second-tone frequency $\omega_{2}$
excited the transitions between the quantum states of the chain, which
resulted in a change of the impedance of this non-linear system. The
chain excitations were detected as a change in the amplitude $|S21|$
and the phase of the microwaves at the probe-tone frequency $\omega_{1}$
(for measurement details, see Supplementary Material 2). 

Below we focus on the most interesting range of magnetic fields close
to full frustration ($\Phi_{R}\approx\Phi_{0}/2$),
where each rhombus represents a $\cos(2\phi)$ Josephson element.
The measurements at the probe-tone frequency (no microwaves at $\omega_{2}$)
show that in this regime the response of the chain to the phase difference
$\varphi$ is indeed periodic with the period $\Delta\varphi=\pi$
(see Supplementary Material 3). 

Figure 3 summarizes the spectroscopic data obtained in the two-tone
measurements. The inset in Fig. 3 shows the resonance frequency $f_{01}$
measured as a function of $n_{g}$ at a fixed magnetic flux $\Phi_{R}=0.5\Phi_{0}$,
$\varphi\approx0$. Note that during the data accuisition
time for the inset in Fig.3 ($\sim1.5$ hours), no long-term shifts
in the offset charge were observed. Also, quasiparticle poisoning
was strongly suppressed due to (a) a larger superconducting gap of
the central island (in comparison with the nearest-neighbor islands)
and a relatively large $E_{C}$ \cite{aumentado2004,bell2012}, and
(b) shielding of the device from stray infra-red photons by the double-wall
light-tight sample holder (see, e.g. \cite{devisser2011}). 

For perfectly symmetric rhombi, the two states corresponding to $\aleph=0,1$
(odd and even number of Cooper pairs on the central island) should
become degenerate ($f_{01}=0$) at $n_{g}=\pm0.5$. Slight asymmetry
of the studied rhombi results in a non-zero $E{}_{01}(n_{g}\!=\!\text{\textpm}0.5)\,\text{\ensuremath{\approx}}\,2E_{1}$
(see also Supplementary Material 4). At $n_{g}=0$ and $E_{2}=0$,
$E_{01}$ is equal to the effective charging energy $E_{C}=15GHz$.
The non-zero energy $E_{2}$ suppresses $E_{01}$ (\ref{eq:E01})
to 6.7 GHz. The energies $E_{1}$,$E_{2}$, and $E_{C}$ were obtained
from the experimental dependences $f_{01}(n_{g})$ measured at different
values of $\varphi$ by fitting them with the computations
of the spectra based on the full Hamiltonian diagonalization. 

As a function of the phase difference across the chain, $E_{2}$ oscillates
with the period $\Delta\varphi=\pi$ (Fig.3).
The dependence $E_{2}(\varphi)$ agrees very well
with the one expected theoretically: $E_{2}(\varphi)=\sqrt{(2E_{2R}\cos(2\varphi))^{2}+(\Delta E_{2R})^{2}}$
(solid red line) where $E_{2R}=4.3$GHz is the energy of an individual
rhombus, ${\Delta}E_{2R}$ is the difference between
these energies for two rhombi. The fit shows that ${\Delta}E_{2R}$
does not exceed $0.1E_{2R}$ for the studied chain. The oscillations
of $E_{2}$ result in a periodic dependence of the measured energy
$E_{01}(\varphi)$. The estimated values of $E_{1}$
do not exceed 0.75 GHz around the optimal values $\varphi=0,\pi,\ldots$
corresponding to a maximum $E_{2}$. This small asymmetry in rhombi
branches is consistent with the reproducibility of submicron junctions
fabricated with the Manhattan-pattern technique \cite{gladchenko,bell2012,bell2012a}.

\begin{figure}
\includegraphics[scale=0.42]{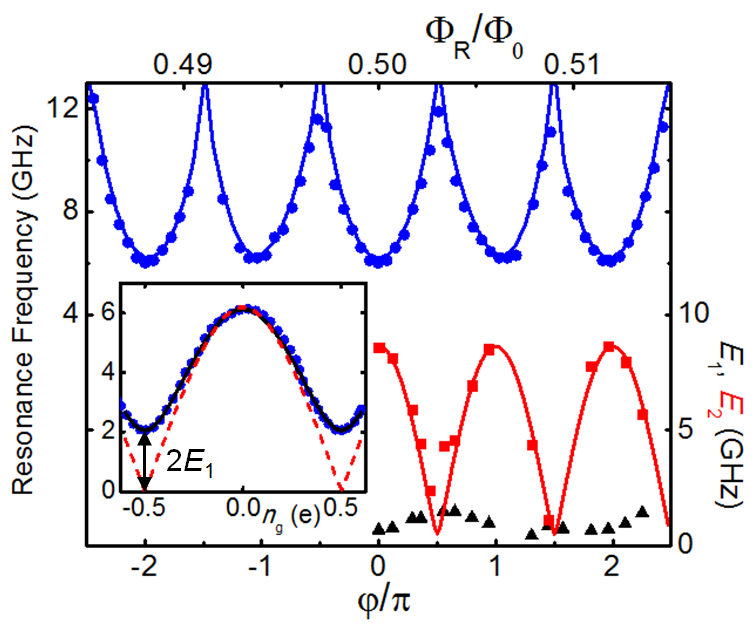}\caption{The resonance frequency $f_{01}$ (blue dots) and energies $E_{1}$
(black triangles) and $E_{2}$ (red squares) versus the phase $\varphi$
across the chain near full frustration $(\Phi_{R}\approx0.5\Phi_{0}$)
at $n_{g}=0$. The theoretical dependences $f_{01}(\varphi)$
(solid blue curve) and $E_{2}(\varphi)$ (solid
red curve) were calculated for $E_{2}=8.5GHz$ and $E_{C}=15GHz$.
The inset: The dependence $f_{01}(n_{g})$ at $\varphi=0$
$(\Phi_{R}=0.5\Phi_{0}$).
The energies $E_{1}$ and $E_{2}$ were extracted from fitting the
dependences $f_{01}(n_{g})$ measured at different $\varphi$
with the calculations based on Hamiltonian diagonalization (solid
black curve). The expected dependence $f_{01}(n_{g})$ for the case
of perfectly symmetric rhombi is shown by the red dashed curve.}
\end{figure}

Because the bandwidth of the second-tone microwave line in our set-up
was limited to $30GHz$, we could not access higher energy states
directly. However, by increasing the second-tone power, we were able
to observe the resonances corresponding to the multiphoton (n=2,3,4)
excitations of the $\left|0\right\rangle \rightarrow\left|2\right\rangle $
transition around the optimal values $\varphi=0,\pi,\ldots.$
The energy $E_{02}$ extracted from these measurements exceeded 50
GHz which is in good agreement with our expectations, and confirms
the validity of the theoretical model (Eq. 1).

The energy relaxation of the state $\left|1\right\rangle $ was measured
by exciting the rhombi chain with a $\pi$-pulse and then measuring
its state after a variable time ${\Delta}t$ (see
Fig.4a inset for pulse sequence). The results of the energy relaxation
measurements at the optimal working point ($\Phi_{R}=\Phi_{0}/2$,
$\varphi=0$) are presented in Fig.4a and Table
1. We also included in Table 1 the $T_{1}$ data for the device with
a large degree of asymmetry (device 2). This asymmetry was caused
by different values of the flux $\Phi_{R}$
in the nominally identical rhombi: variations of the magnetic field
across the chip were caused by conventional (slightly magnetic) microwave
connectors on the sample holder. After replacing these connectors
with the non-magnetic ones, this source of asymmetry was eliminated,
and all the measured devices consistently demonstrated $T_{1}\geq30\mu s$.
The values of $T_{1}$ for the rhombi chains with a high degree of
symmetry are 1-2 orders of magnitude greater than that for less symmetric
rhombi circuits and unprotected Josephson qubits coupled to the same
read-out circuit \cite{bell2012,bell2012a}. The comparison shows
that the symmetry protection supresses the decay rate by almost two
orders of magnitude, in agreement with the large value of the decay
suppression factor $\left(E_{01}/E_{1}\right)^{2}\exp\left(4\sqrt{E_{2}/E_{C}}\right)$
(Table 1). Note that the value of $T_{1}$ at $n_{g}=0$ might be
limited by the Purcell decay into the $LC$ resonator \cite{houck}. 

The quality factor $\omega_{01}T_{1}$ for the protected rhombi chains
exceeds $1\cdot10^{6}$ (see Table 1), and is comparable with that
for the state-of-the-art transmons coupled to 3D cavities \cite{paik2011}
and TiN coplanar resonators \cite{chang2013}. However, the reasons
for such a large $\omega_{01}T_{1}$ in the transmon and in the rhombi
chain are different. In the former case the decay is suppressed by
the carefully designed microwave environment whereas in the latter
case the rhombi symmetry prohibits the decay. 

Table 1.

\begin{tabular}{|c|c|c|c|c|c|c|}
\hline 
Dev. & $n_{g}$ & $\begin{array}{c}
E_{01}\\
GHz
\end{array}$ & $\left(\frac{E_{01}}{E_{1}}\right)^{2}$ & $\exp\left(4\sqrt{\frac{E_{2}}{E_{C}}}\right)$ & $\begin{array}{c}
\omega_{01}T_{1}\\
(10^{6})
\end{array}$ & $\begin{array}{c}
T_{1}\\
(\mu s)
\end{array}$\tabularnewline
\hline 
\hline 
1 & 0 & 6 & 64 & 20 & 1.1 & 30\tabularnewline
\hline 
1 & 0.5 & 2 & 7 & 20 & 0.9 & 70\tabularnewline
\hline 
2 & 0.5 & 4 & 3 & 1 & 0.03 & < 1\tabularnewline
\hline 
\end{tabular}

$\,$

In contrast to the long decay time, the decoherence time in the studied
devices was relatively short ($\sim1\mu s$). The time $T_{2}$ was
determined in Ramsey measurements by applying an $X_{\pi/2}$ pulse
followed by another $X_{\pi/2}$ pulse after a time $\Delta$t
(see Fig.4b inset for pulse sequence). In the spin echo measurements,
a refocusing $X_{\pi}$ pulse was applied between the two $X_{\pi/2}$
pulses. From the Ramsey and spin echo measurements we found $T_{2}=0.45\mu s$
and $T_{echo}=0.8\mu s$. The dephasing time is expected to be long
if $E_{C}\ll E_{2}$. In the studied chain, these energies were of
the same order of magnitude, which resulted in significant dephasing
(Fig.4b). 

\begin{figure}
\includegraphics[scale=0.45]{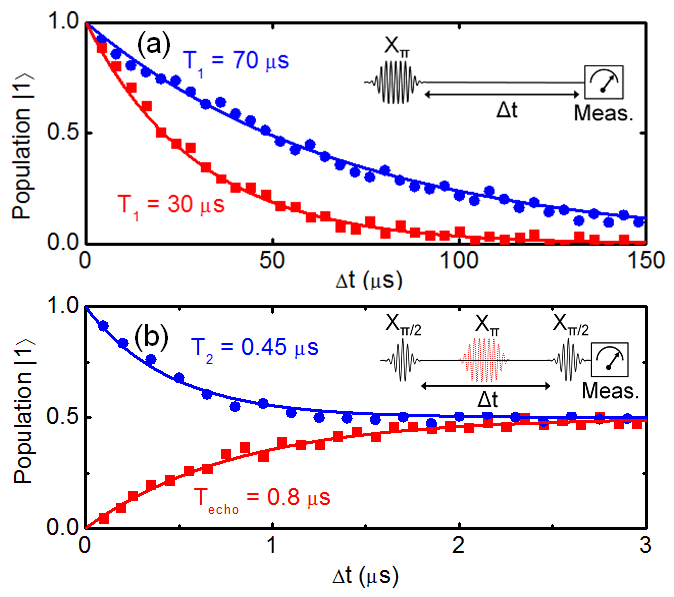}\caption{The time-domain measurements at $\varphi=0$ and
$\Phi_{R}=0.5\Phi_{0}$.
Panel (a): Energy relaxation of the first excited state after application
of a $\pi$-pulse: $n_{g}=0.5$ (blue circles),
$n_{g}=0$ (red squares). Inset: pulse sequence used in the measurement.
Panel (b): Ramsey ($T_{2}$) and spin echo ($T_{echo}$) measurements.
Inset: pulse sequence used in the measurements. For the spin echo
measurement, a refocusing $\pi$-pulse (red
dashed line) was applied at $t=\Delta t/2$.}
\end{figure}

To conclude, we have demonstrated that a Josephson circuit can be
symmetry-protected from the energy decay. We have studied the minimalistic
protected circuit which contains two Josephson rhombi. The symmetry
between the rhombus branches translates into the halving the periodicity
of its Josephson energy $E(\phi)=-E_{2}\cos(2\phi)$, and allows only
the simultaneous transfer of pairs of Cooper pairs on the central
island mutual to both rhombi. The logical states of the protected
qubit correspond to the even and odd number of Cooper pairs on this
island. Our data indicate that the improved design of the Josephson
rhombi \cite{doucot2012} reduces the sensitivity of the chain spectrum
to the offset charge asymmetry. The measured phase and charge dependences
of the energy of the $\left|0\right\rangle \rightarrow\left|1\right\rangle $
transition are in good agreement with our numerical simulations. Symmetry
protection results in a long energy relaxation time $T_{1}$ (up to
70$\mu s$) and a large quality factor $\omega_{01}T_{1}>10^{6}$
of this qubit. The experiments provide a solid foundation for the
next stage \textendash{} the implementation of a qubit with much improved
coherence due to a larger ratio $E_{2}/E_{C}$ which can be realized,
e.g., by parallel connection of a few rhombi chains. 

We would like to thank B. Doucot for helpful discussions. The work
was supported in part by grants from the Templeton Foundation (40381),
the NSF (DMR-1006265), and ARO (W911NF-13-1-0431).

\bibliographystyle{unsrt}

\begin{thebibliography}{10}

\bibitem{Knill05}
Emanuel Knill.
\newblock Quantum computing with realistically noisy devices.
\newblock {\em Nature}, 434:39--44, 2005.

\bibitem{Kitaev20032}
A.Yu. Kitaev.
\newblock Fault-tolerant quantum computation by anyons.
\newblock {\em Annals of Physics}, 303:2 -- 30, 2003.

\bibitem{ioffe2002a}
L.B. Ioffe, M.V. Feigelman, A.~Ioselevich, D.~Ivanov, M.~Troyer, and
  G.~Blatter.
\newblock Topologically protected quantum bits using josephson junction arrays.
\newblock {\em Nature}, 415:503--506, 2002.

\bibitem{Ioffe2002b}
L.~B. Ioffe and M.~V. Feigel'man.
\newblock Possible realization of an ideal quantum computer in josephson
  junction array.
\newblock {\em Phys. Rev. B}, 66:224503, 2002.

\bibitem{doucot2002a}
B.~Doucot and J.~Vidal.
\newblock Pairing of cooper pairs in a fully frustrated josephson-junction
  chain.
\newblock {\em Phys. Rev. Lett.}, 88:227005, 2002.

\bibitem{doucot2003}
B.~Doucot, M.V. Feigelman, and L.B. Ioffe.
\newblock Topological order in the insulating josephson junction array.
\newblock {\em Phys. Rev. Lett.}, 90:107003, 2003.

\bibitem{doucot2005}
B.~Doucot, M.V. Feigelman, L.B. Ioffe, and A.S. Ioselevich.
\newblock Protected qubits and chern-simon theories in josephson junction
  arrays.
\newblock {\em Phys. Rev. B}, 71:024505, 2005.

\bibitem{protopopov2004}
I.V. Protopopov and M.V. Feigelman.
\newblock Anomalous periodicity of supercurrent in long frustrated
  josephson-junction rhombi chains.
\newblock {\em Phys. Rev. B}, 70:184519, 2004.

\bibitem{protopopov2006}
I.V. Protopopov and M.V. Feigelman.
\newblock Coherent transport in josephson-junction rhombi chain with quenched
  disorder.
\newblock {\em Phys. Rev. B}, 74:064516, 2006.

\bibitem{brooks2013}
P.~Brooks, A.~Kitaev, and J.~Preskill.
\newblock Protected gates for superconducting qubits.
\newblock {\em Physical Review A}, 87:052306, 2013.

\bibitem{doucot2012}
B.~Doucot and L.B. Ioffe.
\newblock Physical implementation of protected qubits.
\newblock {\em Rep. Prog. Phys.}, 75:1--20, 2012.

\bibitem{gladchenko}
S.~Gladchenko, D.~Olaya, E.~Dupont-Ferrier, B.~Doucot, L.B. Ioffe, and M.E.
  Gershenson.
\newblock Superconducting nanocircuits for topologically protected qubits.
\newblock {\em Nature Physics}, 5:48--53, 2009.

\bibitem{pop2008}
I.~M. Pop, K.~Hasselbach, O.~Buisson, W.~Guichard, B.~Pannetier, and
  I.~Protopopov.
\newblock Measurement of the current-phase relation in josephson junction
  rhombi chains.
\newblock {\em Phys. Rev. B}, 78:104504, 2008.

\bibitem{pop2012}
I.~M. Pop, B.~Dou\ifmmode~\mbox{\c{c}}\else \c{c}\fi{}ot, L.~Ioffe,
  I.~Protopopov, F.~Lecocq, I.~Matei, O.~Buisson, and W.~Guichard.
\newblock Experimental demonstration of aharonov-casher interference in a
  josephson junction circuit.
\newblock {\em Phys. Rev. B}, 85:094503, 2012.

\bibitem{bell2012}
Matthew~T. Bell, Lev~B. Ioffe, and Michael~E. Gershenson.
\newblock Microwave spectroscopy of a cooper-pair transistor coupled to a
  lumped-element resonator.
\newblock {\em Phys. Rev. B}, 86:144512, 2012.

\bibitem{bell2012a}
M.T. Bell, I.A. Sadovskyy, L.B. Ioffe, A.Yu. Kitaev, and M.E. Gershenson.
\newblock Quantum superinductor with tunable non-linearity.
\newblock {\em Phys. Rev. Lett.}, 109:137003, 2012.

\bibitem{aumentado2004}
J.~Aumentado, Mark~W. Keller, John~M. Martinis, and M.~H. Devoret.
\newblock Nonequilibrium quasiparticles and $2e$ periodicity in
  single-cooper-pair transistors.
\newblock {\em Phys. Rev. Lett.}, 92:066802, 2004.

\bibitem{devisser2011}
P.J. de~Visser, J.J.A. Baselmans, P.~Diener, S.J.C. Yates, A.~Endo, and T.M.
  Klapwijk.
\newblock Number fluctuations of sparse quasiparticles in a superconductor.
\newblock {\em Phys. Rev. Lett}, 106:167004, 2011.

\bibitem{houck}
A.~A. Houck, J.~A. Schreier, B.~R. Johnson, J.~M. Chow, Jens Koch, J.~M.
  Gambetta, D.~I. Schuster, L.~Frunzio, M.~H. Devoret, S.~M. Girvin, and R.~J.
  Schoelkopf.
\newblock Controlling the spontaneous emission of a superconducting transmon
  qubit.
\newblock {\em Phys. Rev. Lett.}, 101:080502, 2008.

\bibitem{paik2011}
Hanhee Paik, D.I. Schuster, L.S. Bishop, G.~Kirchmair, G.~Catelani, A.P. Sears,
  B.R. Johnson, M.J. Reagor, L.~Frunzio, L.I. Glazman, S.M. Girvin, M.H.
  Devoret, and R.J. Schoelkopf.
\newblock Observation of high coherence in josephson junction qubits measured
  in a three-dimensional circuit qed architecture.
\newblock {\em Physical Review Letters}, 107:240501, 2011.

\bibitem{chang2013}
J.B. Chang, M.R. Vissers, A.D. Corcoles, M.~Sandberg, J.~Gao, D.W. Abraham,
  J.M. Chow, J.M. Gambetta, M.B. Rothwell, G.A. Keefe, M.~Steffen, and D.P.
  Pappas.
\newblock Improved superconducting qubit coherence using titanium nitride.
\newblock {\em Appl. Phys. Lett.}, 103:012602, 2013.

\end{thebibliography}

$\,$

\appendix

\section*{SUPPLEMENTARY MATERIALS}

\section*{1. Estimate of the energy difference between protected states and
the matrix element of perturbation. }

\begin{figure}
\begin{centering}
\includegraphics[clip,width=0.9\columnwidth]{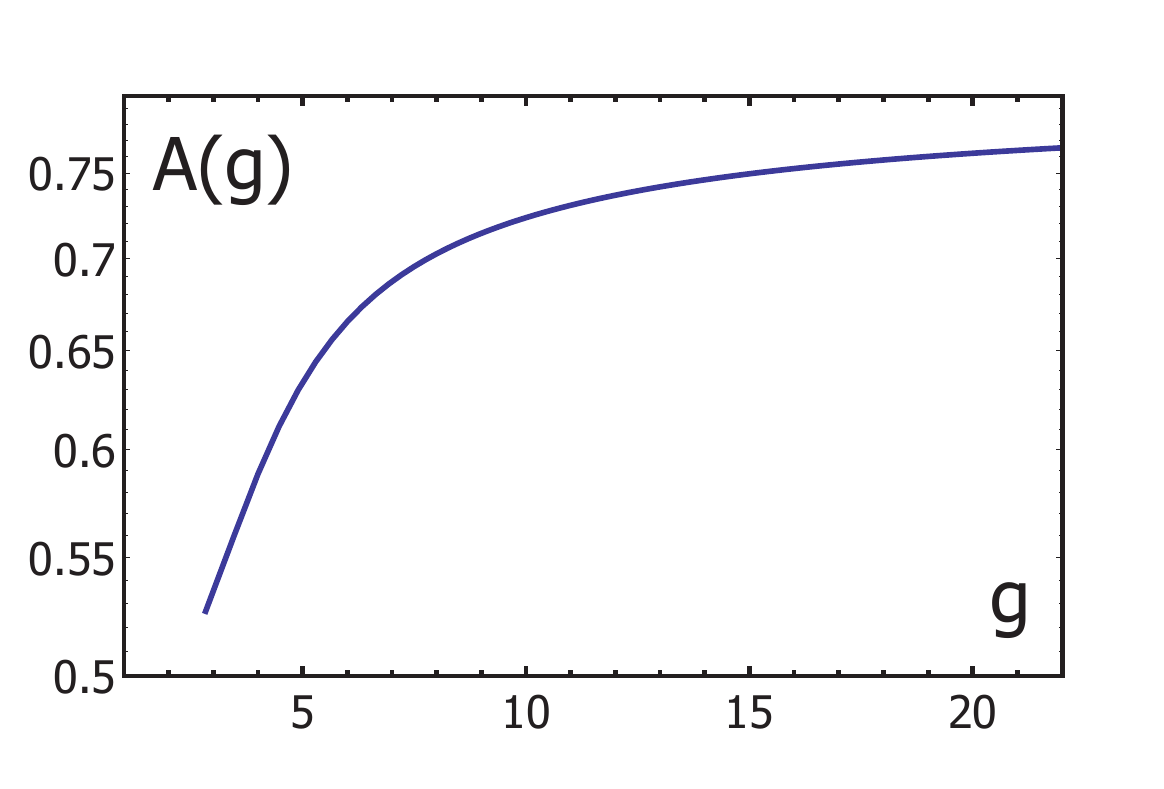}\caption{The dependence of the prefactor $A$ in Eq. (\ref{eq:t}) on $g=4\sqrt{E_{2}/E_{C}}$
. For realistic (not-too-large) values of $g$, $A$ significantly
deviates from its asymptotic value $A(g\gg1)=\sqrt{\frac{2}{\pi}}$
. }

\par\end{centering}

\label{fig:SPrefactor}
\end{figure}

In this section we give the details of the computation of the properties
of the studied system which is equivalent to the discrete oscillator 

\begin{equation}
H_{0}=-2E_{2}\cos(2\phi)+E_{C}(n-n_{g})^{2}.\label{eq:H_0}
\end{equation}
We begin with the ideal system described by (\ref{eq:H_0}). Two lowest
energy states of this problem correspond to wave functions that are
non-zero on even or odd charges as explained in the main text. For
brevity, we shall refer to these states as odd and even states. In
the limit $E_{2}\gg E_{C}$ the splitting of the even and odd states
can be found analytically. In this limit the wave functions in the
phase representation $\Psi(\phi)$ are mostly localized near the minima
of the first term in (\ref{eq:H_0}). Quantum transitions between
two minima result in the coherent mixture of these two states with
the bonding state corresponding to the even and antibonding state
to the odd state in charge representation. To prove the correspondence
between bonding and even states we notice that the bonding state does
not change under transformation $\phi\rightarrow\phi+\pi$, while
the antibonding state changes sign under this transformation. In the
charge basis this implies that bonding state has only even components
while antibonding - only odd ones. In the quasiclassical approximation
the amplitude of this transition in which the phase slips by $\pi$
is

\begin{equation}
t=A(g)g^{1/2}\exp(-g)\omega_{p}.\label{eq:t}
\end{equation}
Here $A(g)\sim1$, $\omega_{p}=4\sqrt{E_{2}E_{C}}$ is the plasma
frequency, the dimensionless parameter $g=4\sqrt{E_{2}/E_{C}}$ has
a physical meaning of the extent of the wave function in the charge
representation: $\left\langle n^{2}\right\rangle =g/4$. In the asymptotic
limit of very large $g\gg1$ the prefactor in (\ref{eq:t}), $A(g)=\sqrt{\frac{2}{\pi}}\approx0.8$
but for moderately large $g$ $A(g)$ is significantly different from
its asymptotic value as shown in Fig. S\ref{fig:SPrefactor}. 

The phase slip transitions by $\pm\pi$ lead to the same states. Adding
these amplitudes we get $2t\cos\pi n_{g}$ that gives energy splitting
between odd and even states 
\begin{equation}
E_{oe}=4t\cos\pi n_{g}\label{eq:E_oe}
\end{equation}

We now discuss the effect of perturbations. Small but non-zero $E_{1}\cos\phi$
term in the Hamiltonian leads to the transition between odd and even
states. In the limit of large $\left\langle n^{2}\right\rangle \gg1$
the amplitude of this transition is $E_{1},$ so in the basis of odd
and even states the Hamiltonian becomes 
\[
H=\left(\begin{array}{cc}
2t\cos\pi n_{g} & E_{1}\\
E_{1} & -2t\cos\pi n_{g}
\end{array}\right)
\]
that gives energy splitting between ground and the first excited state
\begin{equation}
E_{01}=2\sqrt{E_{1}^{2}+\left(2t\cos\pi n_{g}\right)^{2}}\label{eq:E_01}
\end{equation}

We now discuss the effect of external noises. The largest source of
noise is produced by fluctuating potentials that are coupled to the
charge operator. Another smaller but potentially dangerous source
of noise are fluctuations of $E_{1}$ cause by flux fluctuations in
each rhombus: $\delta E_{1}\sim(\delta\Phi/\Phi_{0})E_{J}$ where
$E_{J}$ is the Josephson energy of the small Josephson junction.
These noises are described by the Hamiltonian 
\[
H_{noise}=V(t)n+\delta E_{1}(t)\cos\phi
\]
 In the charge basis the operator $\hat{n}$ is diagonal, so it remains
diagonal in the basis of even and odd states. At $n_{g}=0$ the wave
functions of these states are symmetric in $n$, so the potential
fluctuations are decoupled in the linear order at $n_{g}=0$. For
$n_{g}\neq0$ the diagonal matix elements of $\hat{n}-n_{g}$ can
be found by differetiation of $E_{oe}$:
\[
M=\left(\hat{n}-n_{g}\right)_{oo}-\left(\hat{n}-n_{g}\right)_{ee}=\frac{1}{E_{C}}\frac{dE_{oe}}{dn_{g}}=-\frac{4\pi t}{E_{C}}\sin\pi n_{g}
\]
It contains the same exponential factor as the energy splitting. In
the absence of $E_{1}$ the external potential leads only to the dephasing
proportional to $M^{2}$. For instance, the Johnson noise associated
with the effective impedance $Z$ of the central island leads to dephasing
rate
\begin{eqnarray}
\Gamma_{2}^{Q} & = & \frac{1}{2}\tilde{M}^{2}\sin^{2}\pi n_{g}S_{V}(\omega_{\tau})\label{eq:Gamma_2^Q}\\
\widetilde{M} & = & 4\pi A(g)g^{3/2}\exp(-g)\nonumber \\
S_{V}(\omega) & = & \omega\coth\left(\frac{\omega}{2T}\right)\frac{\Re Z(\omega)}{Z_{Q}}\nonumber 
\end{eqnarray}
Here $Z_{Q}=\hbar/(2e)^{2}$ is the quantum of resistance and $\omega_{\tau}\sim\Gamma_{2}$
is the typical frequency responsible for dephasing process, for realistic
temperatures $\omega_{\tau}\ll T$ so $S_{V}=2T\Re Z/Z_{Q}.$ This
equation assumes slow frequency dependence of the $S_{V}(\omega)$
in a relevant frequency range $\omega\sim\Gamma_{2}$. Small $E_{1}$
results also in a decay:
\[
\Gamma_{1}^{Q}=\frac{1}{2}\tilde{M}^{2}\left(\frac{E_{1}}{E_{01}}\right)^{2}\sin^{2}\pi n_{g}S_{V}(E_{01})
\]

The effect of the flux fluctuations is more direct because they affect
the energy difference between the levels (\ref{eq:E_01}). Assuming
that flux noise is characterized by the $1/f$ spectrum with $\Omega_{0}$
infrared cutoff that translates into the $1/f$ spectrum of $E_{1}$
fluctuations 
\[
\left\langle \delta E_{1}(t)\delta E_{1}(0)\right\rangle _{\omega}=\delta E^{2}/\omega
\]

\[
\Gamma_{2}^{\phi}=\sqrt{\frac{\log(E_{1}/\Omega_{0})}{4\pi}}\frac{E_{1}}{E_{01}}\delta E
\]
 Here $\delta E$ is the typical scale of $E_{1}$ fluctuations caused
by low frequency noise, it is related to the typical flux variations
by $\delta E=\gamma E_{J}(\delta\Phi/\Phi_{0})$ where $E_{J}$ is
the Josephson energy of the small contact in individual rhombus. The
numerical coefficient $\gamma$ in this equation is of the order of
unity in the regime $E_{J}\gg E_{C}$ ($\gamma=\pi)$, similarly to
$E_{2}$ it decreases at $E_{J}\lesssim E_{C}$.

\section*{2. Measurement setup and procedures. }

The main elements of the MW set-up are shown in Fig. S2. The microwave
response of the Josephson rhombi chain coupled to the read-out $LC$
resonator (see Fig. 2) was probed by measuring both the phase and
amplitude of the microwaves traveling along a microstrip feedline
coupled to the resonators. This setup enabled the testing of several
devices fabricated on the same chip in a single cooldown. 

The microwaves at the probe frequency $\omega_{1}$, generated by
a microwave synthesizer, were coupled to the cryostat input line through
a 16 dB coupler and transmitted through the microstrip line coupled
to the $LC$ resonators. The cold attenuators and low-pass filters
in the input microwave lines prevent leakage of thermal radiation
into the resonator. On the output line, two cryogenic Pamtech isolators
($\sim36$ dB isolation between 3 and 10 GHz) anchored to the mixing
chamber attenuate the 5 K noise from the cryogenic HEMT amplifier
(Caltech CITCRYO 1-12, 35 dB gain between 1 and 12 GHz). The amplified
signal is mixed by mixer M1 with the local oscillator signal at frequency
$\omega_{r}$, generated by another synthesizer. The intermediate-frequency
signal $A(t)=A\sin({\Phi}t+\phi)+noise$ at $\Phi\equiv(\omega_{1}-\omega_{r})/2\pi=30MHz$
is digitized by a 1 GS/s digitizing card (AlazarTech ATS9870). The
signal is digitally multiplied by $\sin({\Omega}t)$
and $\cos({\Omega}t)$, averaged over an integer
number of periods, and its amplitude $A$ (proportional to the microwave
amplitude $|S_{21}|$) and phase $\phi$ is extracted as $A=\sqrt{\langle(A(t)sin{\Omega}t)^{2}\rangle +\langle(A(t)cos{\Omega}t)^{2}\rangle }$
and $\phi=\arctan\left[\frac{\langle (A(t)sin{\Omega}t)^{2}\rangle }{\langle (A(t)cos{\Omega}t)^{2}\rangle }\right]$,
respectively. The reference phase $\phi_{0}$ (which randomly changes
when both $\omega_{1}$ and $\omega_{r}$ are varied in measurements)
is found using similar processing of the low-noise signal provided
by mixer M2 and digitized by the second channel of the ADC. This setup
enables accurate measurements of small changes $\phi-\phi_{0}$ unaffected
by the phase jitter between the two synthesizers. The low noise of
this setup allowed us to perform measurements at a microwave excitation
level of -133 dBm which corresponded to a sub-single-photon population
of the tank circuit. In the second-tone measurements, the tested rhombi
chain was excited by the microwaves at frequency $\omega_{2}$ coupled
to the transmission line via a 16 dB coupler. 

In the presence of charge fluctuation on the central island of the
chain, we have applied a sweep-by-sweep averaging method. Since these
fluctuations are slow (on the time scale of seconds), we perform 10-100
full-range sweeps of the second tone frequency (at a rate of 10ms
per one value of $\omega_{2}$) and then perform averaging of the
completed sweeps rather than a point-by-point averaging of the second
tone sweep. 

The sample was mounted inside an rf-tight copper box that provided
the ground plane for the microstrip line and $LC$ resonator. This
box was placed inside another rf-tight copper box in order to eliminate
any stray infrared photons. This nested-box construction was housed
inside a cryogenic magnetic shield (A4k), followed by a superconducting
aluminum magnetic shield. All the connectors inside the magnetic shields
were non-magnetic (EZForm); this reduces the stray magnetic field
at the sample location down to 0.1 mGauss. The entire sample and shielding
construction was anchored to the mixing chamber of a cryogen-free
dilution refrigerator with a base temperature of 20 mK. 

In order to control the charge on the central island of the chain,
a DC gate voltage was applied to the microstrip line via a bias-T.
External noises at the DC port of the bias-T were attenuated with
a voltage divider and a series of low-pass filters and stainless steel
powder filters.

\begin{figure}[h]
\centering{}\includegraphics[clip,width=0.9\columnwidth]{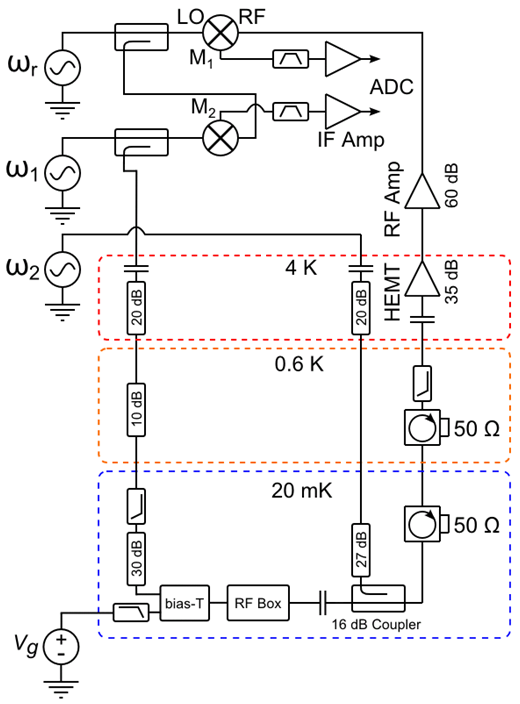}\caption{Simplified circuit diagram of the measurement setup. The microwaves
at the probe frequency $\omega_{1}$ are transmitted through the microstrip
line inductively coupled to the rhombi chain and the $LC$ resonator
located inside the rf sample box. The microwaves at $\omega_{r}$,
after mixing with the microwaves at $\omega_{1}$, provide the reference
phase $\phi_{0}$. The signal at $\omega_{1}$ is amplified, mixed
down to an intermediate frequency $\omega_{1}-\omega_{r}$ by mixer
M2, and digitized by a fast digitizer (ADC). The microwaves at the
second-tone frequency $\omega_{2}$ were used to excite the $\left|0\right\rangle -\left|1\right\rangle $
transitions in the qubit.}
\end{figure}

\section*{3. The probe-tone measurements }

The oscillations in the $LC$ resonance frequency versus the magnetic
flux are shown in Fig. S3 near zero magnetic field (a) and full frustration
$\Phi_{R}=\Phi_{0}/2$ (b). The oscillations in the $LC$ resonance
frequency reflect the dependence of the inductance of the chain in
its ground state as a function of the phase across the chain, $\varphi=2\pi\Phi/\Phi_{0}$.
Near full frustration, the period of oscillations corresponds to $\Delta\varphi=\pi$.
The range of fluxes where the $\pi$-periodicity was observed agrees
well with the theory. The dependence $|S_{21}|(\varphi)$ also demonstrates
avoided crossings observed when the resonance frequency $f_{01}$
coincides with the resonance in the read-out $LC$ resonator (the
\textquotedblleft{}spikes\textquotedblright{} in Fig. S3b), but the
second-tone measurements discussed in the main text provide much more
detailed information on the chain spectrum.

\begin{figure}[H]
\centering{}\includegraphics[clip,width=0.5\columnwidth]{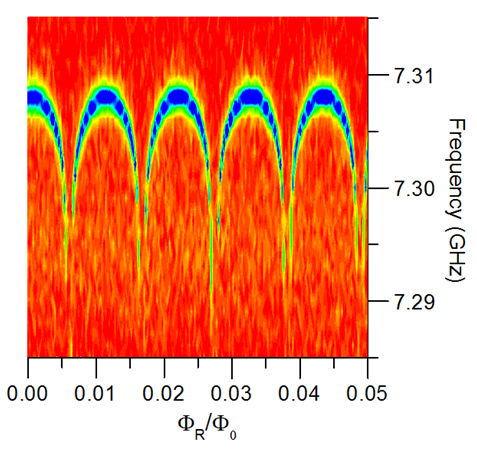}\includegraphics[clip,width=0.5\columnwidth]{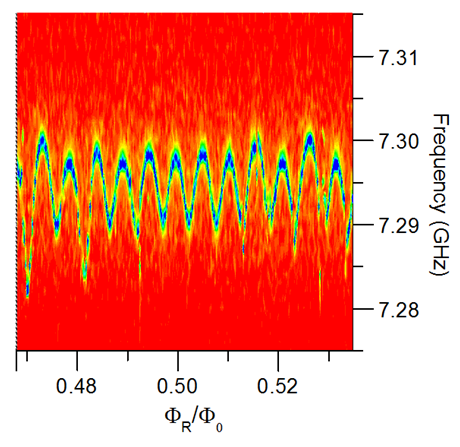}\caption{The color-coded plots show the transmission microwave amplitude $|S_{21}|$
versus the probe-tone microwave frequency and the phase across the
chain near zero field (a) and near full frustration $\Phi_{R}=\Phi_{0}/2$
(b). The period of the inductance oscillations changes from $\Delta\Phi=\Phi_{0}$
($\Delta\varphi=2\pi)$ to $\Delta\Phi=\Phi_{0}/2$$(\Delta\varphi=\pi)$
near full frustration. }
\end{figure}

The energy $E_{2}$ can be estimated from the measurements of the
probe-tone microwave amplitude $|S_{21}|$ versus the magnetic flux
(Fig. S3). The energy $E_{2}$ is inversely proportional to the Josephson
inductance $L_{J}$ of the chain: $E_{2}=(\frac{{\Phi}_{0}}{4{\pi}})^{2}\frac{1}{L_{J}}$
. Both quantities depend on the phase across the chain - $E_{2}(\varphi)=E_{2}\cos(2\varphi)$,$\; L_{J}=L_{J0}/\cos(2\varphi)$
- so $E_{2}\rightarrow0$ and $L_{J}\rightarrow\infty$ when $\varphi\rightarrow\pi/2$.
Using the equivalent circuit in Fig. 2c, one can estimate $L_{J0}$
from the swing of the resonance frequency of the $LC$ resonator coupled
to the chain ($~10$MHz in Fig. S2b). Using $L=3nH$ and $L_{C}=0.4nH$,
we estimate $L_{J0}=19nH$, which translates into $E_{2}=0.39K$ (or
$~7.7$ GHz). This estimate agrees very well with the fitting parameters
used to simulate the chain spectrum.

\section*{4. Spectroscopic data at $n_{g}=0$ and $n_{g}=0.5$}

\begin{figure}[H]
\centering{}\includegraphics[clip,width=0.9\columnwidth]{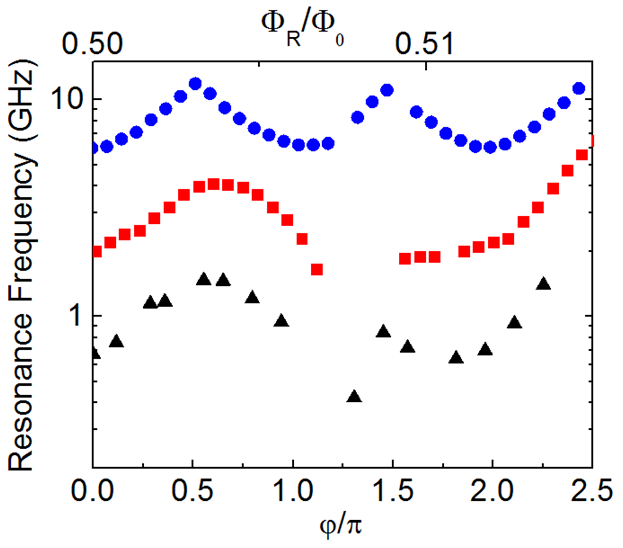}\caption{The dependences of the resonance frequency $f_{01}$ on the phase
$\varphi$ across the chain near full frustration at $n_{g}=0$ (blue
dots) and 0.5 (red squares). For comparison, $E_{1}\left(\varphi\right)$
is shown on the same plot (black triangles). }
\end{figure}

\end{document}